\documentstyle[aps,epsf,preprint]{revtex}

\def\[{\left[}
\def\]{\right]}

\def\({\left(}
\def\){\right)}

\def\d{\delta}
\def\s{\sigma}
\def\l{\lambda}

\def\.{\cdot}

\def\.{\!\cdot\!}
\def\bi{\begin{itemize}}

\def\ei{\end{itemize}}
\def\be{\begin{eqnarray}}
\def\ee{\end{eqnarray}}
\def\bn{\begin{enumerate}}
\def\en{\end{enumerate}}
\def\h{{1\over 2}}

\def\l{\lambda}
\def\p{\partial}
\def\r2{\sqrt{2}}

\def\x{\times}

\def\s{\sigma}

\def\t{\tau}

\def\d{\ \delta}

\def\rr2{{1\over\sqrt{2}}}

\begin{document}

\title{Is There An Ultimate Symmetry in Physics?}
\author{C.S. Lam}
\address{Department of Physics, McGill University\\
3600 University St., Montreal, Q.C., Canada H3A 2T8\\
Email: Lam@physics.mcgill.ca}
\maketitle

\begin{abstract}
A speculative discussion on the roles of various symmetries in physics.
\end{abstract}

\section{The Ever Increasing Symmetry}
This is dedicated to the memory of Bob Sharp, the kindest and the most decent
person in the McGill physics department. Our discussions often turned to
group theory, the subject of his love, with him teaching me.
This talk is in a sense a continuation of those discussions, and I wish Bob
were here to tell me the answer to the question I 
am now posing, namely, whether there
is an ultimate symmetry in physics which can replace all the dynamical laws.

History shows that we became aware of more and more symmetry in Nature
as time went by, and that many important discoveries were
 spear headed by the discovery of a new symmetry. Symmetry
can sometimes be used to replace dynamics, at least partially. 
Thus Newton's first law
in mechanics is a reflection of translational symmetry in space,
and energy conservation is a revelation of the invariance of translation
in time. The huge coriolis force that drives storms on earth into cyclones
and anti-cyclones is caused by the rotational invariance of our physics, 
with conservation of angular momentum being another
manifestation. Einstein's
special relativity is based on Lorentz invariance, and his general relativity
is based on invariance under diffeomorphism. These well-known spacetime
symmetries work from astronomical scale down to the tiniest scale we
know today, around $10^{-18}$ metres. Do they continue to hold down
to much smallest distances? We do not know the answer but we will come 
back to this kind of question later.

Then there is the internal symmetry. Conservation of charge as well
as other additive
quantum numbers can be traced back to the invariance under a phase change,
namely a $U(1)$ symmetry. The discovery of isospin by Heisenberg 
for the nucleon opened up another symmetry, the $SU(2)_F$ symmetry,
with the subscript $F$ standing for `flavour'. 
Unlike the other symmetry we talked about, this symmetry is only approximate,
being broken by the electromagnetic force. 
The subsequent discovery of associated production
led Gell-Mann and Nishijima to postulate strangeness, which in due course
led to the discovery of the $SU(3)_F$ symmetry by Gell-Mann and Ne'eman.
Nowadays we have the Standard Model of strong and electroweak interactions,
whose dynamics is governed by the $SU(3)_c\x SU(2)_L\x U(1)_Y$ symmetry.

\section{Local Symmetry and Dynamics}
According to Noether's theorem, each continuous symmetry leads to
a conserved current $j^\mu(x)$. A local symmetry such as the
gauge symmetry of electromagnetism, or the diffeomorphism symmetry in
Einstein's gravitational equations, contains an infinite number of symmetries,
one at each spacetime point. This means we have an infinite number
of conserved currents. What are they? We surely do not have an infinite
number of conserved quantities in electromagnetism.

To understand this puzzle in the context of a $U(1)$ local
symmetry, let $F^{\mu\nu}(x)$
be the electromagnetic field strength, and $A^\mu(x)$ 
be its vector potential, coupled to a current density $J^\mu(x)$
through the Lagrangian density ${\cal L}(x)=-F^{\mu\nu}F_{\mu\nu}/4+J^\mu 
A_\mu$. Under
an infinitesimal gauge transformation $\d A_\mu(x)=\p_\mu\l(x)$, the
Lagrangian density changes by an amount $\d{\cal L}=J^\mu\p_\mu\l$. 
Using Noether's procedure, we know that the current 
$j^\mu_\l(x)=-F^{\mu\nu}(x)\p_\nu\l(x)$
is driven by a  source $\d{\cal L}$. In other words,
$$\p_\mu j_\l^\mu(x)=\p_\mu\(-F^{\mu\nu}(x)\p_\nu\l(x)\)=\d{\cal L}(x)=
J^\nu(x)\p_\nu\l(x).$$
Since $F^{\mu\nu}$ is antisymmetric in its two indices, 
it can be reduced to
$$\(\p_\mu F^{\mu\nu}(x)+J^\nu(x)\)\p_\nu\l(x)=0,$$
or simply $\p_\mu F^{\mu\nu}(x)+J^\nu(x)=0$ because $\l(x)$ is arbitrary.
This is the Maxwell equations. In other words, local gauge symmetry
gives the full dynamics of Maxwell equations.

What are then the infinite number of conserved charges? 
$j^\mu$ is not conserved unless $J^\mu=0$, so let us consider that case.
The conserved charge is then 
$$Q_\l(t)=\int d^3x j^0_\l(\vec x,t)d^3x=-\int d^3x F^{0k}\p_k\l.$$
Performing an integration by part and assuming as usual that the surface
term vanishes, we come to the realization that $Q_\l(t)=0$ for all $\l(x)$
as long as the Gauss law $\p_kF^{0k}=0$ is obeyed. So the infinite number
of conserved currents simply reduce to the Maxwell equation, and nothing more.

Similar outcome happens in other local symmetries. This suggests that if
we are to have an ultimate symmetry which can replace all dynamics,
local symmetry is likely to be involved.

Local symmetries as we know them in the Standard Model and in Einstein's
gravity are symmetry for the spin-1 gauge fields and the spin-2 gravitational
fields. At this moment we know of no proven local symmetry involving fermions
which can give us dynamical equations of motion for fermions. These would
be necessary if an ultimate symmetry is to be found. Is local supersymmetry
the answer to this prayer?

Even granting that there are local symmetries, or even supersymmetries,
one might still ask what the particular symmetry is and why that particular
symmetry. For that matter why do we live in a four-dimensional spacetime?
Why is $SU(3)_c\x SU(2)_L\x U(1)_Y$ the Standard Model group and not something
else? If there is to be an ultimate symmetry, presumably it should be 
unique and the symmetry itself should be determined somehow. At this moment
the only theory we know of where the symmetry is self-determined is
the string theory. It tells us that the spacetime is really 10 dimensional
and the gauge group, if present, is either $SU(32)$ or $E_8\x E_8$.
Is the string theory pointing us to the direction
of an ultimate symmetry then? We shall come back to discuss this question in
a later section.

If the universe possesses an ultimate symmetry, we are surely
not seeing all of them in this world. That, we know, is because we are living
in a world of low temperature in a late universe, 
where many of the big symmetries are spontaneously
broken. So another criterion for an ultimate symmetry is that it possesses
a mechanism to spontaneously breaking itself at low temperatures, and hopefully
that mechanism is unique as well.

If there is an ultimate symmetry with these characteristics, we are surely
very far from understanding it. String theory or M theory may or may not 
be the answer, but unless their correctness, and the uniqueness of
their symmetry-breaking mechanism can be demonstrated, 
it will not fully satisfy
our criterion here as being the ultimate symmetry.

\section{An Argument for a Supersymmetric Standard Model}
Physics requires three fundamental constants, to tie down the three
independent units we are measuring everything with. It is commonly accepted
that $\hbar$ and $c$ are two of these constants. They turn all units into
powers of energy. The third constant which determines the natural
energy scale is often taken to be the Newtonian gravitational constant $G$,
which gives  a mass scale $M_P\sim 10^{19}$ GeV. 

Accepting $M_P$ to be the natural mass unit, a good question to ask is
why all the particles we know have such tiny masses compared to $M_P$. 
These particles all interact gravitationally.
What keeps them from acquiring a fraction of $M_P$ in their masses? 
In comparison to $M_P$ all the observed masses can be taken to be zero,
and I shall do so in the following.
If you are fussy I will explicitly assume that all the finite masses
we know derive themselves from spontaneously broken symmetry so they are zero 
originally, at high temperature. 

There is a simple mechanism Nature seems to make use of to keep masses zero, 
a mechanism which I shall
refer to as the `spin trick'. A particle with spin $s$ must have
$2s+1$ spin degrees of freedom. The only exception is when the particle 
travels with the speed of light, which means that it is massless. 
So if we can somehow
concoct a theory in which less than $2s+1$ degrees of freedom 
enters, then the particle must be massless. For spin-1
particles, local gauge invariance does just that. It
prevents the particle from having
a physical longitudinal polarization,
thus necessarily it must be massless. Conversely, this may be
taken as an argument for the emergence of gauge theories in the Standard Model. 

For spin-$\h$ fermions, the spin trick demands every fermion to have only
one helicity or chirality for it to be massless. 
This is precisely what the $SU(2)_L\x U(1)_Y$ symmetry
of the Standard Model does. It gives different quantum numbers to different
chiralities of the `same' fermion, making them effectively
two different particles, each having only one chirality. 
Conversely, we can use the massless property of fermions
to argue for the presence of a symmetry group, such as $SU(2)_L\x U(1)_Y$,
which gives different quantum numbers to different chiralities.

What about spin-0 particles? So far, no spinless fundamental particle
has been found, though the Higgs may be looming just around the corner.
Suppose the Higgs is found and it is elementary. It has no spin, so the
spin trick would not work. What keeps its mass small? Well, supersymmetry
might. Supersymmetry links it to a
spin-$\h$ partner, which may be massless because of the spin trick. 
This will force the spinless particle to be massless as well.

We have been dealing with continuous symmetries, but in applying the spin
trick to spin-$\h$ particles, it is important for parity to be violated,
for otherwise there is no way for a particle to have just one helicity
and not two. So parity violation is an integral part of the game plan
as well.

This line of thinking is encouraging for the enterprise of an ultimate
symmetry. The very simple requirement of masslessness ties together
various pieces of physics we know to be true, but taken by themselves
it is hard to see why these pieces must all appear on stage in the
same act. These are local gauge symmetry, the parity-breaking electroweak
symmetry, and possibly supersymmetry as well. Maybe there is after all a
simple grand design where everything occurs for a reason, and not just by
chance? Or is that too optimistic?

\section{An Argument for a String Theory}
The universe started with a big bang, preceeded by a short inflationary
era. What happens {\it before} all that and where 
is the universe going to be put when it is created?
 
The first is an age old question. In the ancient form, it
asks what happens before God created the universe. 
That question led St.~Augustine of Hippo to an important theological
study, published in the book `Confessions'. He essentially concluded that
God is eternal, but He created time when He created the universe
we live in, so it does not make sense to ask
what happens before time is created. He could have said that God created
space to put the universe in as well, but I am not aware of this statement
by him.

Suppose we accept the premise that space and time are created at the 
beginning. Can we construct a theory capable of describing such a creation?
Presumably we need a quantum field theory to do it,
because that is the only theory we know of which deals with the creation
and annihilation of objects.

Ordinary field amplitudes are functions of the spacetime
coordinates $x^\mu$. To be able to create space and time, these
coordinates must be promoted into dynamical fields $X^\mu$, so we need
a replacement of the independent variables which fields depend on.
Denote these new variables as $\s_a$ and $\t$, and call
them the {\it artifical} spacetime coordinates. We know nothing about these
parameters, not even how many $\s_a$'s there are.
However, in this case knowing nothing is knowing a lot, 
because we are saying that these parameters are artificial and 
cannot be observed. Such a theory must involve a large symmetry
as we shall see.

 We encounter a
similar situation in general relativity, where physics in independent
of the choice of spacetime coordinates $x^\mu$. But there
the curvature of spacetime is observable as the gravitational field. 
Nobody has seen artificial gravity, and probably never will, 
so even the curvature of the artificial spacetime cannot be an observable.
How can that be achieved?

Let $n$ be the number of artificial spatial coordinates $\s_a$. If $n=0$,
there is no internal curvature, but it is not a quantum field
theory so we do not have creation and annihilation either. 
If $n=1$, the curvature at
every point in the two-dimensional artificial spacetime has a dimension, 
os it can be affected by a local scale transformation. To make it unobservable, 
we need the theory to be invariant under local scale transformation. A
local scale transformation in two dimensions is a conformal transformation,
so the theory must be diffeomorphic and conformally invariant. This is
the string theory. If $n>2$, the curvature at every point is a tensor.
We do not know of any local invariance which can transform all the
components away. Such a transformation may or may not exist, but if it does
it is certainly not simple because nobody has succeeded in constructing one.
 Barring that as a possibility, we are
led to the string theory as the unique solution of St.~Augustine's premise.
That is amazing, especially because string theory 
demands the (real) spacetime has to be ten dimensional, and the
gauge groups, if present,  to be $SO(32)$ or $E_8\x E_8$.
It is also the only candidate for a consistent quantum gravity theory.
We have no idea whether strng theory is real
physics, but certainly it is beautiful and it goes a long way towards
satisfying the criteria I set out for an ultimate symmetry.

If the argument presented above has any merit, it also suggests that we need
a more intimate connection between cosmology and the formulation of string
theory. I would like to see a string theory which can describe everything,
including the spacetime, and not just the present version where spacetime
is already there for the strings to live in. 
Whether this view is misguided or not
remains to be seen.

We think now that
the five viable superstring theories 
are manifestations of a larger 11 dimensional
`M theory', but we know very little about what that theory is. We do
not even know what symmetry M theory has, though I expect it to be much
larger than the conformal symmetry of the strings. To my mind that is a
very important question to answer.

\section{Is Gravity the Key?}
Can gravity provide the missing link to propel us into an ultimate symmetry?
There is no way to know at this moment, but certainly strong quantum
gravity is a fairly unexplored virgin land which might give us new insights
and bring us new wealth.

For example, if gravity is taken into account,
the maximum number of degrees of freedom $N$ that can be squeezed
into a spherical volume of radius $R$ is proportional to its surface
area $R^2$, but not its volume $R^3$. This 
strange fact comes about in the following way.

Let us put $N$ particles of mass $m$ into the spherical volume.
When the total mass $M=Nm$ reaches $Rc^2/2G$, with $G$ being the Newtonian
gravitational constant, the volume turns into a black hole.
This is the maximum mass the original volume can accommodate,
because adding
more mass will cause the black hole to grow beyond the confines
of the original volume. This maximum $M$ leads to a maximum $N$, because
a particle in a relativistic theory has a natural spread of the order of
its Compton wavelength $\hbar/mc$. This number must be less than $R$
to have the particle stay inside the volume, hence $N=M/m$ 
is bounded by a number of the order of $(R/\ell_P)^2$, 
where $\ell_P=\sqrt{\hbar G/c^3}\sim 10^{-33}$ cm is the Planck length.

Can that strange fact be hinting at us that our physical theory in the
volume can be described by an equivalent theory on the surface of the
sphere, with about one degree of freedom per unit Planck area $\ell_P^2$?
This possibility first pointed out by 't Hooft is known as the
holographic principle. Some examples of that type is known in string
theory but its full implication is perhaps not yet understood. 
Is that an indication for the presence of a larger symmetry to make it
possible?

There are other puzzles. Why is the cosmological constant in the universe
so small? When a black hole evaporates by emitting Hawking radiation, does it
recover all the information that seems to have been lost when matter collapsed
into the black hole? 

We do not know whether these strange facts help us or not in the quest for
an ultimate symmetry, and of course we do not even know that an ultimate
symmetry exists or not.

\end{document}